\documentclass[conference]{IEEEtran}
\IEEEoverridecommandlockouts
\usepackage{booktabs}
\usepackage{float}
\usepackage{cite}
\usepackage{amsmath,amssymb,amsfonts}
\usepackage{algorithmic}
\usepackage{graphicx}
\def\BibTeX{{\rm B\kern-.05em{\sc i\kern-.025em b}\kern-.08em
    T\kern-.1667em\lower.7ex\hbox{E}\kern-.125emX}}
\begin{document}
\title{MVANet: Multi-Stage Video Attention Network for Sound Event Localization and Detection with Source Distance Estimation
}

\author{\IEEEauthorblockN{1\textsuperscript{st} Hengyi Hong}
\IEEEauthorblockA{
\textit{University of Science and}\\
\textit{Technology of China}\\
Hefei, China \\
hyhong@mail.ustc.edu.cn}
\and
\IEEEauthorblockN{\quad\quad\quad\quad\quad\quad\quad\quad\quad 2\textsuperscript{nd} Qing Wang\textsuperscript{*}}\thanks{* Corresponding author}
\IEEEauthorblockA{
\textit{\quad\quad\quad\quad\quad\quad\quad\quad\quad University of Science and}\\
\textit{\quad\quad\quad\quad\quad\quad\quad\quad\quad Technology of China}\\
\quad\quad\quad\quad\quad\quad\quad\quad\quad Hefei, China \\
\quad\quad\quad\quad\quad\quad\quad\quad\quad qingwang2@ustc.edu.cn}
\and
\IEEEauthorblockN{\quad\quad\quad\quad\quad\quad\quad\quad\quad 3\textsuperscript{rd} Jun Du}
\IEEEauthorblockA{
\textit{\quad\quad\quad\quad\quad\quad\quad\quad\quad University of Science and}\\
\textit{\quad\quad\quad\quad\quad\quad\quad\quad\quad Technology of China}\\
\quad\quad\quad\quad\quad\quad\quad\quad\quad Hefei, China \\
\quad\quad\quad\quad\quad\quad\quad\quad\quad jundu@ustc.edu.cn}
\and
\IEEEauthorblockN{\quad 4\textsuperscript{th} Ruoyu Wei}
\IEEEauthorblockA{
\textit{\quad iFlytek Research}\\
\quad Hefei, China \\
\quad rywei@iflytek.com}
\and

\IEEEauthorblockN{\quad\quad\quad\quad\quad\quad\quad\quad\quad\quad\quad\quad 5\textsuperscript{th} Mingqi Cai}
\IEEEauthorblockA{
\textit{\quad\quad\quad\quad\quad\quad\quad\quad\quad\quad\quad\quad iFlytek Research}\\
\quad\quad\quad\quad\quad\quad\quad\quad\quad\quad\quad\quad Hefei, China \\
\quad\quad\quad\quad\quad\quad\quad\quad\quad\quad\quad\quad mqcai@iflytek.com}
\and
\IEEEauthorblockN{\quad\quad\quad\quad\quad\quad\quad\quad\quad\quad\quad\quad 6\textsuperscript{th} Xin Fang}
\IEEEauthorblockA{
\textit{\quad\quad\quad\quad\quad\quad\quad\quad\quad\quad\quad\quad iFlytek Research}\\
\quad\quad\quad\quad\quad\quad\quad\quad\quad\quad\quad\quad Hefei, China \\
\quad\quad\quad\quad\quad\quad\quad\quad\quad\quad\quad\quad xinfang@iflytek.com}
}

\maketitle

\begin{abstract}
Sound event localization and detection with source distance estimation (3D SELD) involves not only identifying the sound category and its direction-of-arrival (DOA) but also predicting the source’s distance, aiming to provide full information about the sound position. This paper proposes a multi-stage  video attention network (MVANet) for audio-visual (AV) 3D SELD. Multi-stage audio features are used to adaptively capture the spatial information of sound sources in videos. We propose a novel output representation that combines the DOA with distance of sound sources by calculating the real Cartesian coordinates to address the newly introduced source distance estimation (SDE) task in the Detection and Classification of Acoustic Scenes and Events (DCASE) 2024 Challenge. We also employ a variety of effective data augmentation and pre-training methods. Experimental results on the STARSS23 dataset have proven the effectiveness of our proposed MVANet. By integrating the aforementioned techniques, our system outperforms the top-ranked method we used in the AV 3D SELD task of the DCASE 2024 Challenge without model ensemble. The code will be made publicly available in the future.
\end{abstract}


\section{Introduction}

Sound event localization and detection (SELD) is capable of predicting the direction-of-arrival (DOA) of active sound sources over time and performing sound event detection (SED) \cite{b1}. It has a wide range of applications, such as virtual reality, audio surveillance, and industrial inspection \cite{b1-2}. 

Recently, joint modeling of SED and DOA estimation is becoming a major research focus. SELDnet, proposed by Adavanne et al. \cite{b2}, employed two parallel branches to perform SED and DOA estimation, respectively. Shimada et al. introduced an activity-coupled Cartesian DOA (ACCDOA) method \cite{b3}. Cao et al. proposed an Event-Independent Network V2 (EINV2) \cite{b4}. Shimada et al. subsequently extended ACCDOA to Multi-ACCDOA \cite{b5} for distinguishing and localizing multiple overlapping sound events of the same category. Kim et al. proposed an angular-distance-based YOLO approach \cite{b8-2} for the SELD task, able to address the polyphony problem. The SELD task of the DCASE 2024 Challenge introduced the source distance estimation (SDE) task, significantly increasing the complexity by extending two-dimensional (2D) SELD to three-dimensional (3D) SELD. Research on 3D SELD is increasingly growing, specifically, \cite{b5-2} proposed an extension of the multi-ACCDOA method to include distance information. \cite{b5-3} focused on estimating the distance and DOA of binaural sound sources in scenarios with a walking listener, using motion-based cues. \cite{b5-4} used a convolutional recurrent neural network (CRNN) with an attention module for continuous distance estimation from single-channel audio signals.

\begin{figure*}[t]
\centerline{\includegraphics[width=\linewidth,height=0.260\textheight]{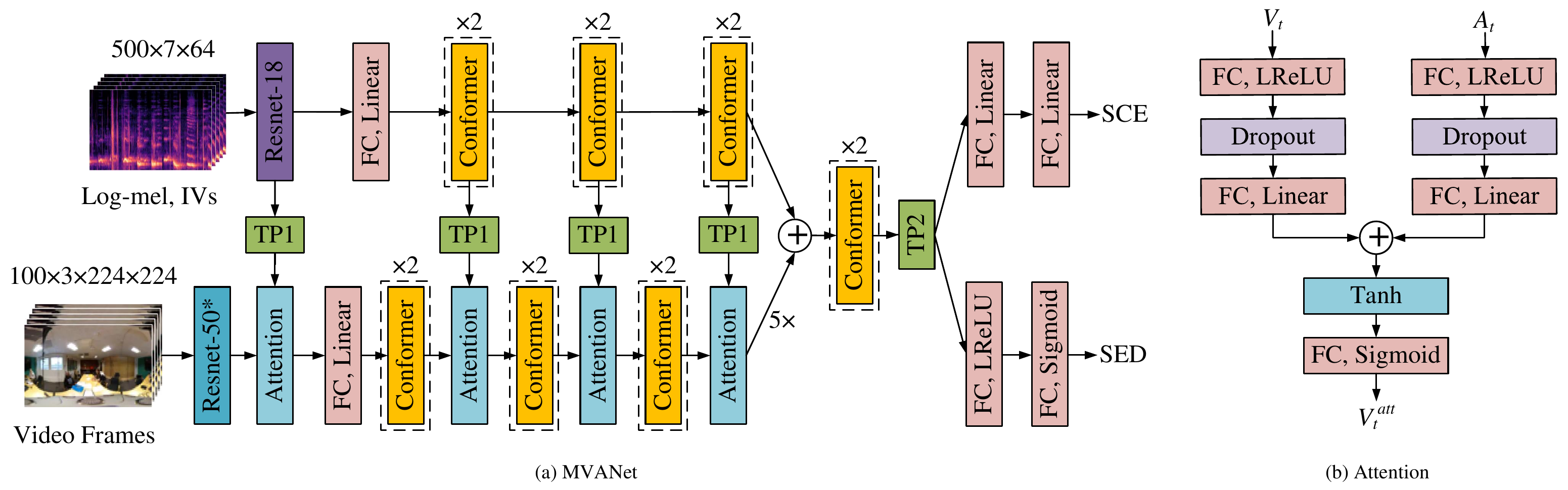}}
\caption{An illustration of the MVANet architecture and the details of the attention module. * denotes frozen pre-trained parameters. ``TP1'' and ``TP2'' refer to the average and max pooling along the temporal dimension, respectively. ``5×'' denotes repetition five times over time. $\textcircled{+}$ represents the summation operation.}
\label{models}
\end{figure*}

Humans often rely on multi-modal senses, especially hearing and vision, to perceive and understand their surroundings. In the audio-visual (AV) SELD tasks, video can offer high spatial precision, while audio can detect obscured objects. Therefore, several research focused on integrating audio and visual information to achieve better SELD performance. Current approaches to multi-modal fusion in SELD generally fall into feature-level fusion and decision-level fusion. Feature-level fusion maps audio and visual features into a unified feature space for subsequent processing \cite{b6,b7}. Decision-level fusion utilizes visual information through face/object bounding boxes or human keypoint detection \cite{b7,b8}. Among these fusion methods, audio and video data are processed independently before integration. However, the information embedded in video frames can be excessively redundant, such as containing background elements, similar objects, and other information unrelated to the sound events. Therefore, AV SELD methods often failed to surpass the performance of models trained using only audio data \cite{b7-2,b7-3,b6}.

Audio-guided video attention was first introduced and successfully applied to audio-visual event localization in \cite{b9}. Inspired by this work, we propose a multi-stage video attention network (MVANet) to solve the AV 3D SELD task. Audio embeddings from multiple network layers are used to enhance visual embeddings by dynamically focusing on spatial information related to sound events, leveraging the complementary characteristics between audio and video modalities. The main contributions of this paper are fourfold: (1) We propose a multi-stage attention network, aiming at extracting audio-related information in video frames ; (2) We propose an novel output format that simultaneously performs sound event detection and source coordinate estimation (SED-SCE), combining DOA estimation with distance estimation tasks; (3) We employ effective data augmentation techniques such as audio-visual pixel swapping and audio-visual data synthesis, along with audio pre-trained models; (4) By integrating the above three key technologies, our system significantly outperforms the baseline and also surpasses the top-ranked method we used in the AV 3D SELD task of the DCASE 2024 Challenge.

\section{Proposed Method}
We propose a novel multi-stage video attention network (MVANet) designed to enhance the visual features related to sound source locations in video frames through audio features, as shown in Fig. 1a. We chose ResNet-18 \cite{b16} as the audio encoder. Meanwhile, we utilized the ResNet-50 \cite{b10}, pre-trained on ImageNet, as the visual encoder. Audio and visual features from the same stage are aligned to ensure temporal synchronization. Subsequently, the aligned features are fed into the attention module to enhance the characteristics of the sound source locations in video frames, as shown in Fig. 1b. Furthermore, we employ the Conformer to extract both local and global contextual features from the input sequence \cite{b10-2}. We leverage audio embeddings from four stages to fully exploit source location-related information in video frames. Finally, our model performs SED and SCE through two parallel branches, each consisting of two fully-connected (FC) layers. The SED branch is responsible for identifying the types of sound sources, while the SCE branch estimates their 3D spatial locations. This dual-branch structure allows our model to effectively handle 3D SELD tasks. In the following sections, we will elaborate on the key techniques of our method.

\subsection{Attention}
Attention mechanisms have shown excellent performance in many multi-modal learning tasks, such as audio-visual sound separation \cite{b10-3} and audio-visual event localization \cite{b10-4}. In this paper, we employ the attention modules at four stages of the network to enhance audio-visual fusion. The details of the attention module are shown in Fig. 1b. Specifically, we define the attention function as ``Att'', which can adaptively learn from the visual feature vector \({\mathbf{V}}_t \in {\mathbb{R}^{k \times 1}}\) and the audio feature vector \({\mathbf{A}}_t \in {\mathbb{R}^{n \times 1}}\) at each time step \(t\). The output \({\mathbf{V}}_t^{\rm att}\) is given by:
\begin{equation}
{\mathbf{V}}_t^{\rm att}={\rm Att}({\mathbf{V}}_t,{\mathbf{A}}_t)\label{eq1},
\end{equation}
where \({\mathbf{V}}_t^{\rm att} \in {\mathbb{R}^{k \times 1}}\) is a vector of attention weights, the specific calculation process is as follows:
\begin{equation}
{\mathbf{A}_t} = {\mathbf{W}_a}({\sigma_1}({\mathbf{U}_a}({\mathbf{A}_t})))\label{eq2},
\end{equation}
\begin{equation}
{\mathbf{V}_t} = {\mathbf{W}_v}({\sigma_1}({\mathbf{U}_v}({\mathbf{V}_t})))\label{eq3},
\end{equation}
\begin{equation}
\mathbf{V}_t^{\rm att} = {\sigma_3}({\mathbf{W}_{av}}{\sigma_2}({\mathbf{A}_t} + {\mathbf{V}_t}))\label{eq4},
\end{equation}
where \({\mathbf{U}_a} \in {\mathbb{R}^{n \times n}}\), \({\mathbf{U}_v} \in {\mathbb{R}^{k \times k}}\), \(\mathbf{W}_{av} \in {\mathbb{R}^{k \times d}}\) are linear projection parameters. \({\mathbf{W}_a} \in {\mathbb{R}^{d \times n}}\) and \({\mathbf{W}_v} \in {\mathbb{R}^{d \times k}}\) are two transformation functions that project audio-visual features onto the same feature dimension \(d\). For simplicity, we have omitted the bias of the linear layers in these formulae. \({\sigma_1}(\cdot)\), \({\sigma_2}(\cdot)\) and \({\sigma_3}(\cdot)\)  represent the Leaky ReLU function, the hyperbolic tangent function, and the sigmoid function, respectively. By repeatedly applying the attention mechanism at different layers of the network, spatial information related to sound sources in video frames is fully exploited.

\subsection{SED and SCE}
To address the 3D SELD task, we propose integrating source distance information into the corresponding DOA estimation. We multiply the normalized Cartesian coordinates of the sound event, i.e., the DOA, by the source distance to obtain the true source Cartesian coordinates. The output format aims to predict the true Cartesian coordinates of active sound sources, where the direction of the coordinate vector represents the DOA, and the length of the coordinate vector represents the source distance. The SCE branch uses linear output to encompass the range of DOA and distance values. A multi-task learning framework is used to address SED and SCE tasks with the loss function expressed as:
\begin{equation}
\mathcal{L} = {\beta _1}\mathcal{L}_1 + {\beta _2}\mathcal{L}_2,\label{eq5}
\end{equation}
where \({\beta _1}\) = 1 and \({\beta _2}\) = 2 are the weights for the SED loss \(\mathcal{L}_1\) and the DOA loss \(\mathcal{L}_2\). Binary cross-entropy (BCE) is used as the loss function for the SED branch, and mean squared error (MSE) is used as the loss function for the SCE branch, defined as follows:
\begin{align}
\mathcal{L}_1 =&  - \frac{1}{{CT}}\sum\limits_{t,c} {[{y_{t,c}}\log {{\hat y}_{t,c}}} + (1 - {y_{t,c}})\log (1 - {{\hat y}_{t,c}})],\label{eq6} \\
\mathcal{L}_2 =& \frac{1}{{CT}}\sum\limits_{t,c} {\left\| {(\mathbf{{\hat o}}_{t,c} - {\mathbf{o}_{t,c}}){y_{t,c}}} \right\|^2},\label{eq7}
\end{align}
where \(\{ {\hat y_{t,c}},{y_{t,c}}\}\) and \(\{ \mathbf{{\hat o}}_{t,c},{\mathbf{o}_{t,c}}\}\) represent the model outputs and ground truth of the \(c\)-th sound event at the \(t\)-th frame for SED and SCE, respectively. \(T\) and \(C\) denote the number of frames in a batch and sound event classes, respectively.

\subsection{Video Data Augmentation and Pre-Training}
The STARSS23 dataset contains approximately 3.8 hours of audio-visual training data \cite{b11}, which is insufficient for training a robust 3D SELD model. To acquire more audio-visual data, we employ two effective data augmentation methods. The first method utilizes the audio-video simulation method proposed by \cite{b12,b15-2}. Spatialized sound events are generated using room impulse responses (RIR) from the METU-SPARG \cite{b13} and TAU \cite{b13-2} datasets. A spatial audio synthesizer extracts audio from YouTube videos and convolves it with RIRs. This method provided us with approximately 6 hours of audio and video data. 
The STARSS23 dataset contains simultaneous $360^{\circ}$ video recordings with a resolution of $1920 \times 960$, corresponding to an azimuth angle range of $\left[180^{\circ},-180^{\circ}\right]$ and an elevation angle range of $\left[-90^{\circ}, 90^{\circ}\right]$. We extend our audio channel swapping (ACS) \cite{b14} to the audio-visual pixel swapping (AVPS) method to augment the amount of audio-visual data \cite{b6}. Unlike our previous work \cite{b7-1}, We generate new video frames by flipping and rotating the original frames \cite{b6}. With these two methods, we obtained approximately 80 hours of audio and video data. In order to obtain the audio pre-trained model, we simulate 40 hours of multi-channel audio using the Spatial Scaper library \cite{b15-2}, and augment the data by a factor of 7 using ACS. The parameters of the audio-visual model were initialized with the pre-trained audio-only ResNet-Conformer (RC) model \cite{b16}.

\section{EXPERIMENTS}

\subsection{Implementation Details}

We evaluated the AV 3D SELD task using the official development set of the DCASE 2024 Challenge. The audio was recorded in First-Order Ambisonics (FOA) format with a sampling rate of 24kHz. Additionally, the video was captured using a panoramic camera at a resolution of $1920 \times 960$ and a frame rate of 29.97 fps \cite{b11}. Visual features were extracted using a pre-trained ResNet-50 network at a frame rate of 10 fps. Global average pooling was applied at the last layer of the ResNet-50 to obtain a $7 \times 7$ feature map, which was then reshaped into a 49-dimensional feature vector. Since the length of the audio-visual clips input to the network is fixed at 10 seconds, the shape of the visual features is $100 \times 49$. It is noteworthy that before being input to the Conformer, the output of the visual encoder is fed into a linear layer to achieve a 256-dimensional video embedding. As for audio data, we applied short-term Fourier transform (STFT) with a hop length of 20 msec to extract 4-channel log-mel spectrograms and 3-channel acoustic intensity vectors (IVs). 

The experiments employed the Adam optimizer \cite{b17-2} with a tri-stage learning rate scheduler \cite{b17}, with an upper limit of the learning rate set at 0.0005. Additionally, we adopted our early fusion ResNet-Conformer, which won first place in the DCASE 2024 Task 3 \cite{b18} for fair comparison. Specifically, we repeated the $7 \times 7$ visual feature map for each frame five times to align the visual features with the audio features along the temporal dimension. The obtained visual features were concatenated with audio features, resulting in fused audio-visual features with a shape of $7 \times 500 \times 71$. Finally, the audio-visual features were fed into the ResNet-Conformer network for training. In order to evaluate the SELD performance, we utilized the F-score (\({F_{20^\circ/1 }}\)) related to the spatial location and category of the sound source, the DOA error (\(DOAE\)) related to the angular distance error, and the relative distance error (\(RDE\)) related to the distance error.

\subsection{Ablation Studies}
Firstly, we investigate the 3D SELD performance of the MVANet when trained from scratch, and the results are presented in Table~\ref{tab:1}. The model named ``Wang \cite{b18}'' denotes our previous method that uses early fusion for audio-visual data and ranks the first place in DCASE 2024 Challenge Task 3. As shown in the first three rows of Table~\ref{tab:1}, we utilized the \textit{dev-set-train} from the STARSS23 dataset as the AV basic data for training. Our proposed MVANet significantly outperforms the baseline with a 79.6\% improvement in \({F_{20^\circ/1 }}\), a 32.0\% improvement in \({DOAE}\), and a 23.9\% improvement in \({RDE}\). The MVANet also surpasses our previous method \cite{b18} across three key metrics. Specifically, the performance of MVANet in \({F_{20^\circ }}\) yields a 15.3\% improvement. This demonstrates the effectiveness of multi-stage attention for 3D SELD by leveraging the complementarity between audio and video data.
\begin{table}[t]
\caption{Performance comparison of different models with several audio-visual data augmentation setups on the \textit{dev-set-test} of STARSS23 dataset. `BASE': AV basic data, `ENH1': AV basic data + AVPS, ‘ENH2’: AV basic data + AV simulated data + AVPS, ‘ENH3’: AO basic data + AO simulated data + ACS.}
\begin{center}
\begin{tabular}{c|c|c|c|c}
\toprule
Model & Setup & \textbf{\({F_{20^\circ/1 }}\)↑} & \textbf{\(DOAE\)↓} & \textbf{\(RDE\)↓} \\
\midrule
Baseline & BASE & 0.113 & 38.40° & 0.360 \\
Wang \cite{b18} & BASE & 0.176 & 28.92° & 0.296 \\
MVANet & BASE & \textbf{0.203} & \textbf{26.11°} & \textbf{0.274} \\
\midrule
Wang \cite{b18} & ENH1 & 0.309 & 17.93° & 0.274 \\
MVANet & ENH1 & \textbf{0.328} & \textbf{17.44°} & \textbf{0.260} \\
\midrule
Wang \cite{b18} & ENH2 & 0.441 & 15.58° & 0.251 \\
MVANet & ENH2 & \textbf{0.458} & \textbf{14.61°} & \textbf{0.250} \\
\midrule
AO-RC & ENH3 & 0.439 & 15.98° & 0.262 \\
\bottomrule
\end{tabular}
\label{tab:1}
\end{center}
\end{table}

In addition, we also evaluate the performance of the proposed MVANet when applying audio-visual data augmentation. To this end, we apply the AVPS method step by step on the AV basic data of STARSS23 and on the AV data including simulation, creating two distinct AV data augmentation setups, denoted as ``ENH1'' and ``ENH2''. For training the audio-only (AO) RC model, we apply the ACS method to the AO basic data of STARSS23 dataset and simulated data, forming an AO data augmentation configuration, labeled as ``ENH3''. By comparing the results under two data configurations (``ENH1'' and ``ENH2''), it is observed that the proposed MVANet achieves consistent improvement in SELD performance over our previous methods denoted ``Wang \cite{b18}''. Specifically, under the ``EHN2'' setup, the \(DOAE\) of MVANet was approximately 6.2\% lower than that of Wang \cite{b18}. Furthermore, the proposed MVANet yields superior performance on all of the three evaluation metrics compared to the AO-RC model shown in the last two rows, both of which are trained with the same amount of data. 

\subsection{Comparison with Other Competitive Methods}

We compare our proposed MVANet with the top three competitive AV models \cite{b18,b19,b20} in the DCASE 2024 Challenge Task 3, all of which adopt a parameter initialization strategy. This strategy has been proven to be effective. To compare with the other competitive methods, we initialize the parameters of audio branch in MVANet with the RC model trained using audio data, while the video branch and the attention module are randomly initialized. Berghi \cite{b19} and Li \cite{b20} achieve similar SELD results as shown in the second and third rows of Tabel~\ref{tab:2}. 
It is worth noting that Berghi \cite{b19} and Li \cite{b20} did not generate simulated audio-visual data to augment the training set; therefore, we train a MVANet using the same AV data as Berghi \cite{b19} and Li \cite{b20} with the results presented in the fourth row of Table~\ref{tab:2}. MVANet exhibits significant advantages across all three metrics compared to Berghi \cite{b19} and Li \cite{b20} under the same training setup. Specifically, the \({F_{20^\circ/1 }}\) of the MVANet is 22.8\% higher than that of Berghi \cite{b19}, the \(DOAE\) is 22.8\% lower than that of Berghi \cite{b19}, and the \(RDE\) is 18.3\% lower than that of Berghi \cite{b19}. 

Another comparison is between our proposed MVANet and our previous method, as shown in the first and last rows. MVANet outperforms our previous method in \cite{b18}, with \({F_{20^\circ/1 }}\) increasing from 0.550 to 0.567, \(DOAE\) decreasing from 13.16$^\circ$ to 12.37$^\circ$, and \(RDE\) decreasing from 0.250 to 0.239. Additionally, as shown in the first row, the \(RDE\) of our best single system in the DCASE 2024 Challenge was only 0.250, which was due to our use of the pre-trained model optimized for SED and DOA estimation tasks in DCASE 2023 Challenge Task 3, lacking of pre-training for source distance estimation. By comparing the best results in Table~\ref{tab:1} with those in the last row of Table~\ref{tab:2}, it is evident that employing an audio pre-trained model for parameter initialization significantly enhances performance in the 3D SELD task. 

\begin{table}[t]
\caption{Performance comparison of the proposed MVANet with 3D SELD models of other teams on the dev-set-test of the STARSS23 dataset. `PT': loading pre-trained audio models.}
\centering
\begin{tabular}{c|c|c|c|c}
\toprule
Model & Setup & \textbf{\({F_{20^\circ/1 }}\)↑} & \textbf{\(DOAE\)↓} & \textbf{\(RDE\)↓} \\
\midrule
Wang \cite{b18}& ENH2 + PT & \textbf{0.550} & \textbf{13.16°} & \textbf{0.250} \\
Berghi \cite{b19} & ENH1 + PT & 0.403 & 18.00° & 0.300 \\
Li \cite{b20} & ENH1 + PT & 0.392 & 18.70° & 0.310 \\
\midrule
MVANet & ENH1 + PT & 0.495 & 13.90° & 0.245 \\
MVANet & ENH2 + PT & \textbf{0.567} & \textbf{12.37°} & \textbf{0.239} \\
\bottomrule
\end{tabular}
\label{tab:2}
\end{table}

\begin{figure}[t]
\centerline{\includegraphics[width=0.95\linewidth,height=0.19\textheight]{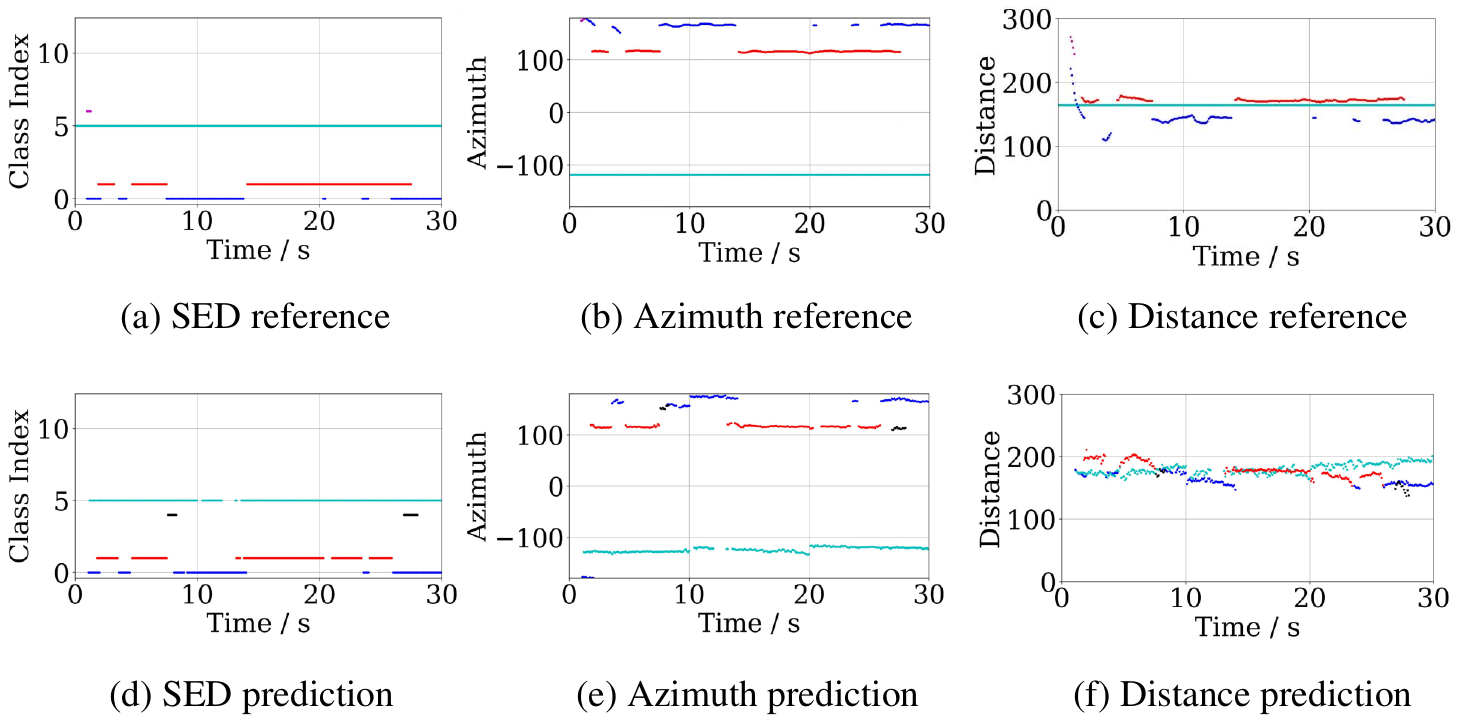}}
\caption{Visualization comparison of the results under `ENH2 + PT' with ground truth. Different colors represent different source categories.}
\label{6_p}
\end{figure}
Fig.~\ref{6_p} visually demonstrates that the 3D SELD results of our proposed MVANet under the ``ENH2 + PT'' training setup are close to the ground truth. However, there is still room for improvement in SDE performance.

\section{CONCLUSION}

In this paper, we propose the MVANet, which aims to adaptively capture the spatial information related to sound sources in video frames using multi-stage audio features. We also introduce a novel source Cartesian coordinate output format, combining the DOA with distance estimation of sound sources. By employing two effective data augmentation techniques and a pre-training strategy, the proposed model outperforms our previous top-ranking method in the AV 3D SELD task of the DCASE 2024 Challenge. In the future, we will explore more effective multi-modal fusion strategies to further improve the AV 3D SELD performance. 

\vfill
\pagebreak

\end{document}